\documentclass[prd,floatfix,nofootinbib,preprintnumbers,showpacs]{revtex4}
\usepackage{epsfig}
\usepackage{graphicx}
\usepackage{amsfonts}

\def\beq{\begin{equation}}
\def\eeq{\end{equation}}
\def\be{\begin{eqnarray}}
\def\ee{\end{eqnarray}}
\def\nnb{\nonumber}
\def\ed{\end{document}}

\newcommand{\gsim}{\lower.7ex\hbox{$\;\stackrel{\textstyle>}{\sim}\;$}}
\newcommand{\lsim}{\lower.7ex\hbox{$\;\stackrel{\textstyle<}{\sim}\;$}}

\begin{document}

\begin{center}
{\LARGE\bf
Direct CP violation of $B \to l \nu$ in unparticle physics
}

\vspace{0.2cm}
Chao-Shang Huang$^{a,b}$,~  Xiao-Hong Wu$^c$\\ 
\vspace{0.2cm}
$a$ Department of Physics, Henan Normal University,
Xinxiang, Henan 453007, China\\
        $^b$ ITP, Academia Sinica, P.O. Box 2735, 100080 Beijing, China \\
        $^c$ School of Physics, Korea Institute for Advanced Study,
         207-43, Cheongryangri 2-dong, Dongdaemun-gu, Seoul 130-722, Korea
\end{center}
\begin{abstract}
We have investigated the effects of unparticles in decays $B \to l
\nu$. It is found that the direct CP violation in the decays,
which is zero in SM, can show up due to the CP conserving phase
intrinsic in unparticle physics. For $l=\tau$, the direct CP
asymmetry can reach $30\%$ for the scalar unparticle contribution,
and $100\%$ for the longitudinal vector unparticle contribution
under the constraints of ${\rm Br}(B\to\tau\nu)$ and $\nu e$
elastic scattering. If both unparticle-lepton coupling
universality and unparticle-quark coupling universality are
assumed the constraint from $Br(\pi\to \mu\nu)$ leads that the
direct CP violation in $B\to l\nu$ can only reach at most $8\%$
and $1\%$ for scalar and vector unparticle contributions
respectively if $d_{\cal U} < 2$. If the direct CP violation is
observed in the future it would give strong evidence for the
existence of unparticle stuff.
\end{abstract}
\pacs{12.60.-i, 13.25.Hw, 11.30.Er}
\maketitle

The scale invariant and conformal invariant quantum field theories
in two and higher dimensions have been extensively
investigated~\cite{sca}. For example, it was noted that there is an
infra-red fixed point in SU(N) gauge theory with massless fermions
if the number of fermion representations $f=\frac{11}{2}N-k$ with
$k<<N$~\cite{gross}. They have been widely applied in a number of
fields in physics. However, unfortunately, particle physics at the
low energy (say, the electro-weak scale) can not be described by
scale invariant theories because $[P_{\mu}, D]=i P_{\mu}$, where D
is the scale transformation generator, leads to that the mass
spectrum of scale invariant system is either continuous or
massless. Therefore, scale invariant stuff, if it exists,
is made of unparticles.

Georgi recently proposed an interesting suggestion to determine
experimentally whether unparticle stuff actually
exists~\cite{Georgi:2007ek}. Assuming there are interacting two sectors
above a high energy scale $M_{\cal U}$: one is the SM, the other is a
nontrivial scale invariant renormalizable quantum field system of
scale dimension $d_{\cal U}$, the non-renormalizable terms due to the
interactions between the two sectors exist below $M_{\cal U}$ and, in the
form of operator product, can generally be written as
\be
\label{int} O_{SM}O_{SI}/M_{\cal U}^n~~~~{\rm with}~~n>0,
\ee
where $O_{SM}$ and
$O_{SI}$ are an operator with dimension $d_{SM}$ built of SM
fields and an operator with dimension $d_{SI}$ built of fields in
the scale invariant sector respectively. What can we see below a
low energy scale $\Lambda$ which is much smaller than $M_{\cal U}$?
Can we see the effects of the nontrivial scale invariant sector below
$\Lambda$? For this purpose Georgi presented to use effective
field theory approach. In the effective field theory below
$\Lambda$ the scale invariant operators $O_{SI}$ match onto the
unparticle operators and the interactions (\ref{int}) match onto
interactions of the form
\be
\label{intl} \frac{C_{\cal U}
\Lambda^{d_{SI}-d_{\cal U}}}{M_{\cal U}^n}O_{SM}O_{\cal U}
\ee
where $d_{\cal U}$ is the
scaling dimension of the unparticle operator $O_{\cal U}$ and $C_{\cal U}$ is a
constant coefficient function.

There are a lot of papers which discuss phenomenological
consequences in unparticle physics~\cite{Cheung:2007ue,
Luo:2007bq, Chen:2007vv, Ding:2007bm, Liao:2007bx, Aliev:2007qw,
Li:2007by, Duraisamy:2007aw, Lu:2007mx, Stephanov:2007ry,
Fox:2007sy, Greiner:2007hr, Davoudiasl:2007jr, Choudhury:2007js,
Chen:2007qr, Aliev:2007gr, Mathews:2007hr, Zhou:2007zq,
Ding:2007zw, Chen:2007je, Liao:2007ic, Bander:2007nd,
Rizzo:2007xr, Cheung:2007ap, Goldberg:2007tt, Chen:2007zy,
Kikuchi:2007qd}. As it is well-known, there are many theoretical
models beyond SM. The most important issue is how to discriminate
one from the others. For unparticles, there is a distinguishing
feature which differs from all other well-known models. That is,
there are CP conserving phases in unparticle propagators. Because
the phases appear in propagators of unparticles and conserve CP
they should act as the strong phases in physics processes. The
phases would show up in various processes. An interesting
observation by Chen and Geng is to measure the direct CP violation
in rare leptonic decays $B\to l^+l^-$~\cite{Chen:2007vv}. Because
the direct CP violation in the decays is zero in SM and all other
well-known models, its discovery would give strong evidence for
the existence of unparticle stuff. However, from the experimental
point of view, it is far away to do the measurements because the
branching ratios of rare leptonic B decays have not been determined
yet. In contrary, leptonic charged B decays have been observed and
the branching ratio in $B^+\to \tau^+ \nu_{\tau}$ has been recently
given by the BaBar collaboration~\cite{btau1}, \be {\rm Br}(B^+\to
\tau^+\nu) &=& (0.9\pm 0.6({\rm stat.})\pm 0.1({\rm syst.}))
\times 10^{-4} \\
\nnb &<& 1.7\times 10^{-4} \;\; {\rm at}\;\; 90\% \;\;{\rm C.L.}
\label{data1} \ee The experimental data of Br of $B^-\to
\tau^-\nu$ by the Belle collaboration is~\cite{btau2} \be {\rm
Br}(B^-\to \tau^-\nu) = (1.79^{+0.56}_{-0.49}({\rm
stat.})^{+0.46}_{-0.51}({\rm syst.}))\times 10^{-4}. \ee It is
expected to measure the direct CP violation of $B^\pm \to \tau^\pm
\nu_{\tau}$ in B factories and super B factories~\cite{supb} in
the near future. Because there are no hadronic final states in the
leptonic charged B decays and consequently the direct CP violation
in the decays is zero in SM and all other well-known models, its
discovery would give strong evidence for the existence of the
unparticle stuff. In this paper we shall study the phase effects
of unparticles in the leptonic charged B decays. We show that the
direct CP violation in $B^+\to \tau^+ \nu_{\tau}$ can reach $30\%$
for scalar unparticle contribution, and $100\%$ for longitudinal
vector unparticle contribution under the constraints of ${\rm
Br}(B\to\tau\nu)$ and $\nu e$ elastic scattering in the reasonable
region of parameters in unparticle physics.

In discussions so far, the unparticle operator $O_{\cal U}$ is
assumed to be the SM singlet. However, it is not unreasonable to
assume there are SM non-singlet unparticle operators. It was
argued in ref.~\cite{sei1} that for every $\frac{3}{2}N_c<N_f<3
N_c$ there is a non-trivial fixed point in super Yang-Mills
theories with the gauge group $SU(N_c)$ and $N_f$ quark flavors in
the fundamental representation of the gauge group. Therefore, for
this range of $N_f$, the infrared theory is a non-trivial four
dimensional superconformal field theory and the charged scale
invariant fields could exist. As pointed out in ref~\cite{sei2},
although the results are obtained for supersymmetric field
theories, the insights obtained are expected to be also applicable
for non-supersymmetric theories, at least at a qualitative level,
because the dynamical mechanisms explored are standard to gauge
theories. The charged currents in SM would interact with the
charged scale invariant fields~\cite{sei1,hort} by exchanging some
charged particles and/or charged scale invariant fields at the
high scale. Therefore, we can parameterize the effective couplings
of unparticles to the charged leptonic currents as follows. \be
\frac{1}{\Lambda^{d_{\cal U}-1}}\sum_{i=e,\mu,\tau} C_V^{l_i}
\bar{l}_i\gamma_{\mu}(1-\gamma_5)\nu_i O_{\cal U}^\mu
+\frac{1}{\Lambda^{d_{\cal U}}}\sum_{i=e,\mu,\tau}
C_S^{l_i}\bar{l}_i\gamma_{\mu}(1-\gamma_5)\nu_i
\partial^\mu O_{\cal U} + h.c. , \label{lep}
\ee where $O_{\cal U}$ and $O_{\cal U}^\mu$ are scalar and vector
operators respectively. In principle the vector and axial-vector
couplings may be different and there may be scalar and
pseudo-scalar couplings.
For simplicity here we assume the above left-handed couplings. 
Similarly we parameterize the effective
couplings of unparticle to the charged quark currents as \be
\frac{1}{\Lambda^{d_{\cal
U}-1}}\sum_{q^\prime=u,c,t,q=d,s,b}C_V^{q^\prime
q}\bar{q}^\prime\gamma_{\mu}(1-\gamma_5)q O_{\cal U}^\mu +
\frac{1}{\Lambda^{d_{\cal
U}}}\sum_{q^\prime=u,c,t,q=d,s,b}C_S^{q^\prime
q}\bar{q}^\prime\gamma_{\mu}(1-\gamma_5)q\partial^\mu O_{\cal U} +
h.c. \label{qua} \ee In eqs.(\ref{lep}) and (\ref{qua}), C's are
dimensionless parameters\footnote{Scale factors have been included
into the definitions of C's, e.g., $C_S^l=c^l_S
\Lambda^d_{SI}/M_{\cal U}^n$.} which have been assumed
to be real for simplicity.
Recently the direct CP violation in
leptonic B decays due to charged unparticle effects is also analyzed
by introducing scalar and pseudo-scalar couplings of unparticles to
quarks and leptons~\cite{Zwicky:2007vv}.

Due to the interaction between the charged currents in SM and the
charged scale invariant fields which could give a correction to
the fixed point of the scale invariant sector, the scale
(conformal) invariance of the scale (conformal) invariant sector
would be violated. Because the unparticles, which we consider in
the paper, are electromagnetically charged, it is expected that
breaking of scale (conformal) invariance would be weaker than that
for the colored unparticles. As analyzed in
refs.~\cite{Fox:2007sy,Bander:2007nd}, even for the SM singlet
scalar unparticle the scale invariance would be violated after
electroweak symmetry breaking if it couples to Higgs sector, and
the scale, at which the conformal invariance is broken by the
Higgs vacuum expectation value, depends on the coupling of SM
singlet scalar unparticle to Higgs and the conformal window can
extend down to the scale of B meson mass if the coupling is small
enough (say, smaller than $0.01$). We do not have detailed models
to describe the scale invariant sector and do not know the details
of how the scale invariance is broken. However, we know that
arising of the scale invariance breaking must be at a low energy.
So we shall parameterize our ignorance with an infra-red cutoff
scale $\mu$ for the unparticle propagator. That is, we modify the
propagators of scalar and vector particles in
refs.~\cite{Georgi:2007si,Cheung:2007ue} to
\be \Delta_{O_{\cal U}}= i \frac{A_{d_{\cal U}}}{2\sin(d_{\cal
U}\pi)}
 (P^2-\mu^2)^{d_{\cal U}-2}e^{-i\phi_{\cal U}},
\Delta_{O_{\cal U}^\mu}^{\mu\nu}= i \frac{A_{d_{\cal
U}}}{2\sin(d_{\cal U}\pi)}
 \Pi^{\mu\nu}(P^2-\mu^2)^{d_{\cal U}-2}e^{-i\phi_{\cal U}},
\label{lon} \ee where the constant $\phi_{\cal U}=(d_{\cal
U}-2)\pi$ and \be \label{ten} \Pi^{\mu\nu}=-g^{\mu\nu}+P^\mu
P^\nu/P^2 \ee if assuming the vector unparticle is transverse,
$\partial_\mu O_{\cal U}^\mu=0$. As pointed out in
ref.~\cite{berg}, the vector unparticle scale dimension $d_{\cal
U}< 2$ is allowed when scale invariance is broken at a scale
$\mu\ge 1 GeV$. Furthermore, it is shown that the decay into an
unparticle has a non-integrable singularity in the decay rate for
$d_{\cal U}< 1$~\cite{Georgi:2007ek}. Therefore, we shall assume
$1 < d_{\cal U} < 2$ in numerical calculations below.

We find the transverse vector unparticle does not contribute to
the decay $B^\pm\to l^\pm \nu$. By a straightforward calculation
we obtain the contributions of scalar unparticles to the amplitude
in $B^\pm\to l^\pm \nu$, \be A^U=f_B C_S^lC_S^{bu}\frac{A_{d_{\cal
U}}}{2\sin(d_{\cal U}\pi)} \frac{m_B^{2(d_{\cal
U}-1)}}{\Lambda^{2d_{\cal U}}}(1-\frac{\mu^2}{m_B^2})^{d_{\cal
U}-2} e^{-i\phi_{\cal U}} \bar{\nu} \not\!{P} (1-\gamma_5)l,
\label{aunp} \ee
where the decay constant $f_B$ is defined as usual, \be
<0|\bar{b}\gamma_\mu\gamma_5 u|B^+(P)>=-if_B P_\mu\,. \ee
In eq. (\ref{aunp}), $A_{d_{\cal U}}$ is the normalization factor of phase
space for unparticle stuff.
For the phenomenological analysis, one can constrain the product
$C_S^l C_S^{bu}$, so that besides
the infra-red cutoff $\mu$, we have three new parameters: $d_{\cal
U}$, $\Lambda$, $C_S^l C_S^{bu}$. The amplitude in SM is \be
A^{SM}=f_B \frac{g^2}{8m_W^2}V^*_{ub}\bar{\nu}
\not\!{P}(1-\gamma_5)l \,. \label{asm} \ee So the branching ratio in
$B^+\to l^+ \nu$ is, by using eqs. (\ref{aunp}), (\ref{asm}), \be
\label{br} {\rm Br}(B^+\to\l^+\nu)&=&\tau_B\frac{p_c}{8\pi
m_B^2}\Sigma_{\lambda}|A^U+A^{SM}|^2
\\ \nnb &=& {\rm Br} (B^+\to\l^+\nu)^{SM}|1+re^{-i(\phi_w+\phi_{\cal U})}|^2
\ee where \be {\rm Br}(B^+\to\l^+\nu)^{SM}=\frac{G_F^2 m_B
m_{l}^2}{8\pi}(1-\frac{m_{l}^2}{m_B^2})^2\tau_B f_B^2|V_{ub}|^2
\ee is the branching ratio in SM, \be \label{r}
r=\frac{8C_S^lC_S^{bu}}{g^2|V_{ub}|}\frac{A_{d_{\cal U}}}
{2\sin(d_{\cal U}\pi)}\frac{(m_B^2)^{d_{\cal
U}-1}m_W^2}{\Lambda^{2d_{\cal U}}}(1-\frac{\mu^2}{m_B^2})^{d_{\cal
U}-2} \ee with $V_{ub}=|V_{ub}|e^{-i\phi_w}$
and $\phi_w$ is the angle $\gamma$ of the unitarity triangle,
$\gamma = 77^{+30^\circ}_{-32^\circ}$ (CKMfitter) ~\cite{CKMfitter}
and $\gamma= 88 \pm 16^\circ$ (UTfit) ~\cite{UTfit},
which we use $\gamma=80^\circ$ in the numerical studies.
We define the Br of
$B \to l \nu$ as the average of ${\rm Br}(B^-\to\l^-\bar{\nu})$
and ${\rm Br}(B^+\to\l^+\nu)$. In eq.(\ref{br}), $\phi_w$ is the
weak phase from the CKM matrix element and
the phase $\phi_{\cal U}$ comes from the propagator of unparticle
and is CP conserved, like the strong phase in QCD. As emphasized
in ref.~\cite{Georgi:2007si,Cheung:2007ue}, the existence of
$\phi_{\cal U}$ is an unusual feature in unparticle physics. A
non-zero $\phi_{\cal U}$ leads to direct CP violation in
$B^\pm\to\l^\pm\nu$: \be A_{\rm CP}&\equiv &
\frac{\Gamma(B^-\to\l^-\bar{\nu}) - \Gamma(B^+\to\l^+\nu)}{
      \Gamma(B^-\to\l^-\bar{\nu}) + \Gamma(B^+\to\l^+\nu)}\\
\nnb &=&
\frac{2r\sin\phi_{\cal U}\sin\phi_w}{1+r^2+2r\cos\phi_{\cal U}\cos\phi_w}
\label{cpv}
\ee

In the above calculations we have assumed the vector unparticle is
transverse, like the vector particles in SM. However, it could
contain the longitudinal component because we do not know its
properties at present. We now discuss the contribution of the
longitudinal component to the branching ratio and direct CP
violation in the decays $B^\pm\to l^\pm \nu$. Setting
$\Pi_{\mu\nu}=P_\mu P_\nu/P^2$ in eq. (\ref{lon}), we obtain \be
\label{vec} A^U_{\rm
long}=\frac{C_V^lC_V^{bu}}{C_S^lC_S^{bu}}\frac{\Lambda^2}{m_B^2}
A^U \ee where $A^U$ is the contribution of unparticles to the
decay amplitude due to scalar unparticles, eq. (\ref{aunp}).
Therefore, the contribution of longitudinal component dominates
that of scalar unparticle by a factor of $(\Lambda/m_B)^2$ if the
coupling constants of vector unparticle are in the same order as
those of scalar unparticle in magnitude. The Br of $B^\pm\to l^\pm
\nu$ will constrain the parameters, $C_V^lC_V^{bu}$ or $\Lambda$
and $d_{\cal U}$ strongly, which we shall discuss below.

\begin{figure}
\includegraphics[width=8cm] {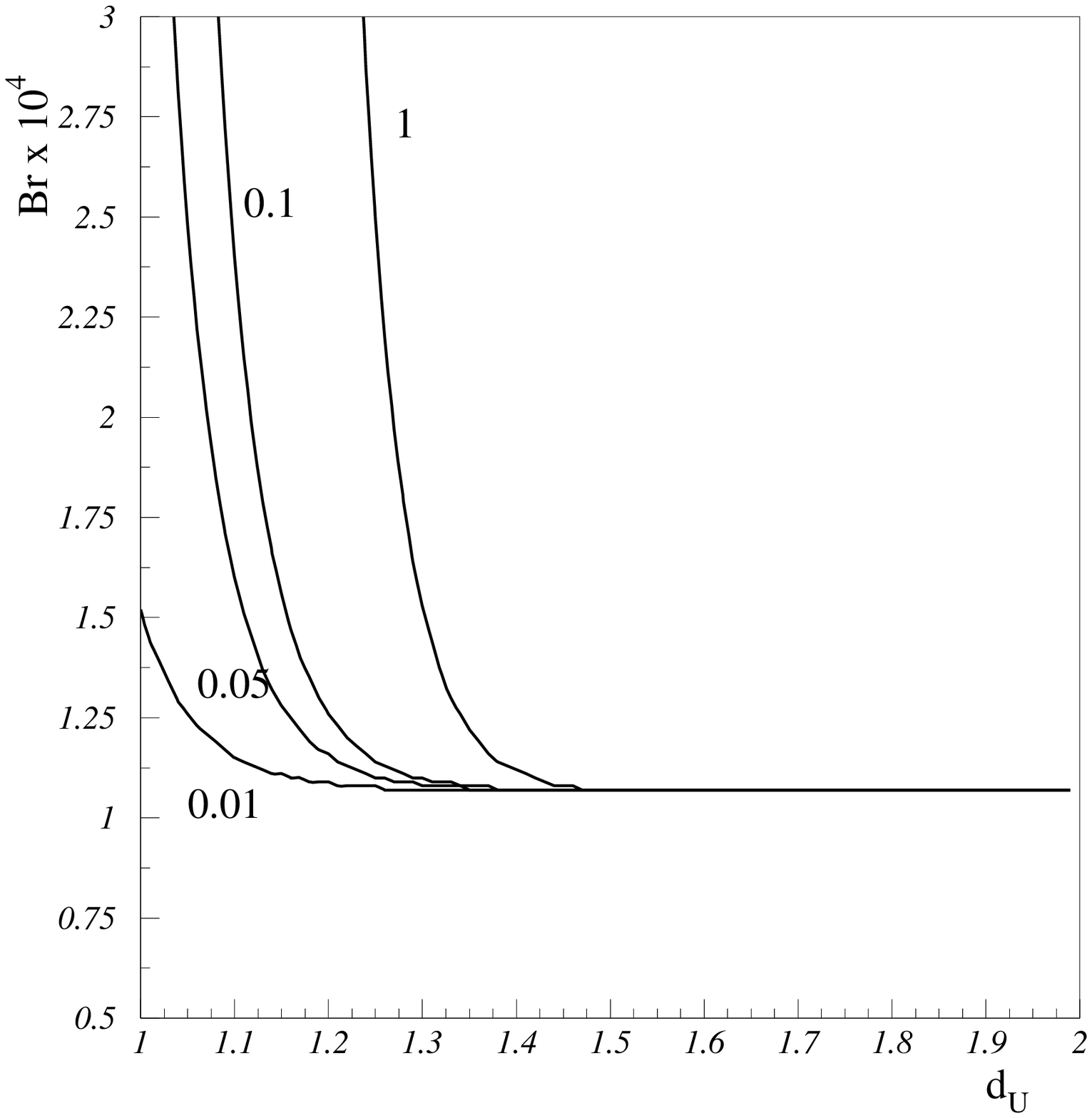}
\includegraphics[width=8cm] {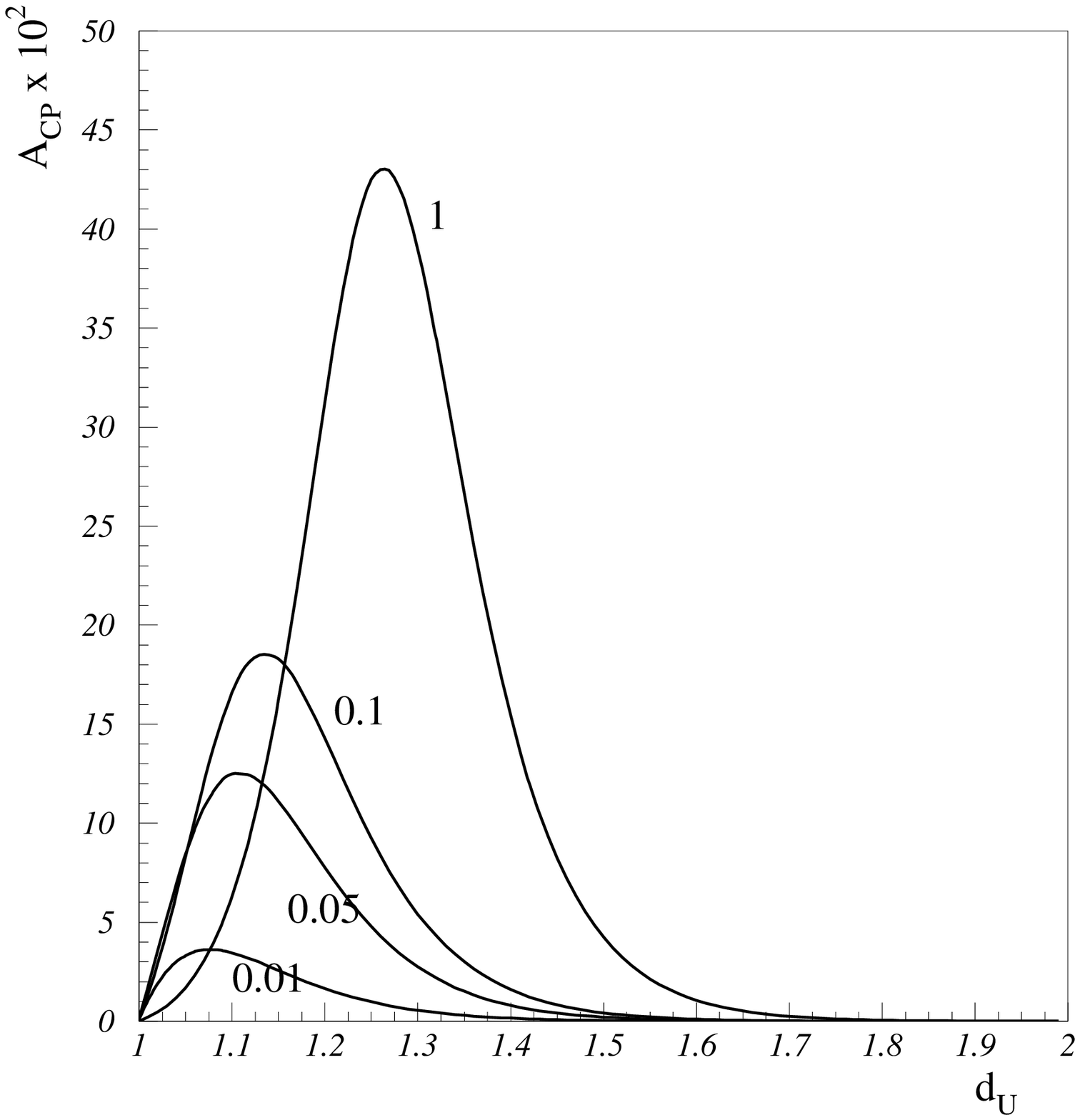}
\caption{ \label{scalar} Scalar unparticle contribution to Br and
$A_{\rm CP}$ of $B^- \to \tau \bar{\nu}$. From right to left, the
corresponding coefficients are $C^{bu}_S C^l_S = 1.$, $0.1$,
$0.05$, $0.01$. }
\end{figure}

In numerical calculations we fix  $\Lambda = 1$ TeV
and first consider the case of scalar unparticle.
In Fig.  (\ref{scalar}), we plot the dependence of Br and $A_{\rm
CP}$ as a function of scaling dimension $d_{\cal U}$, for
different multiplicity of $C^{bu}_S C^l_S = 1.0$, $0.1$, $0.05$,
$0.01$, with $\mu=0$. For $C^{bu}_S C^l_S = 0.01$, the unparticle
contribution is quite small. However, in the case of $C^{bu}_S
C^l_S = 1$, the contribution is large, the Br exceeds experimental
bound when $d_{\cal U} \le 1.3$, and $A_{\rm CP}$ can reach $30\%$
under the Br constraint. We estimate the effects of the infra-red
cutoff by taking $\mu=0.5 m_B$ and the effects are quit small.

\begin{figure}
\includegraphics[width=8cm] {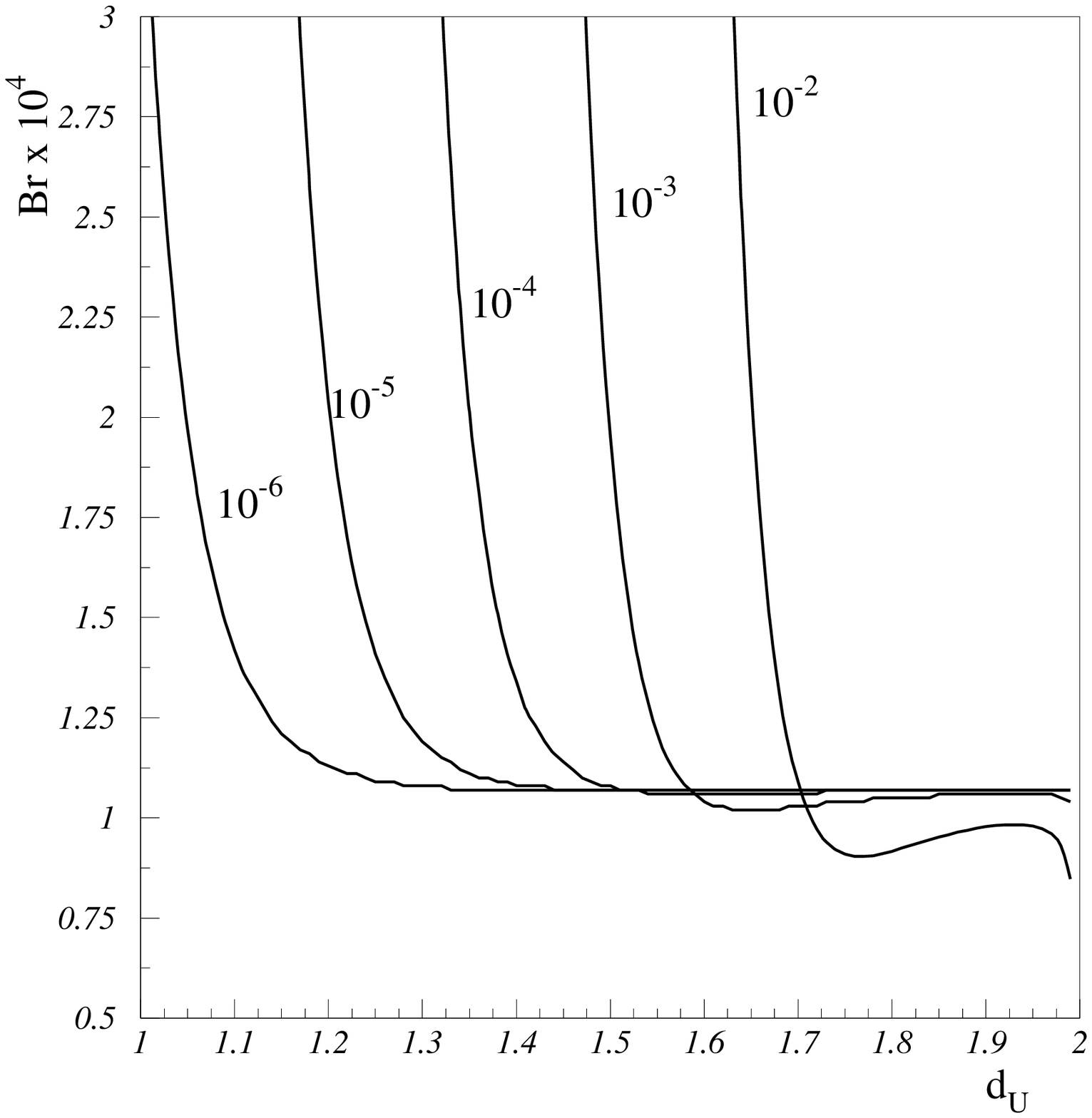}
\includegraphics[width=8cm] {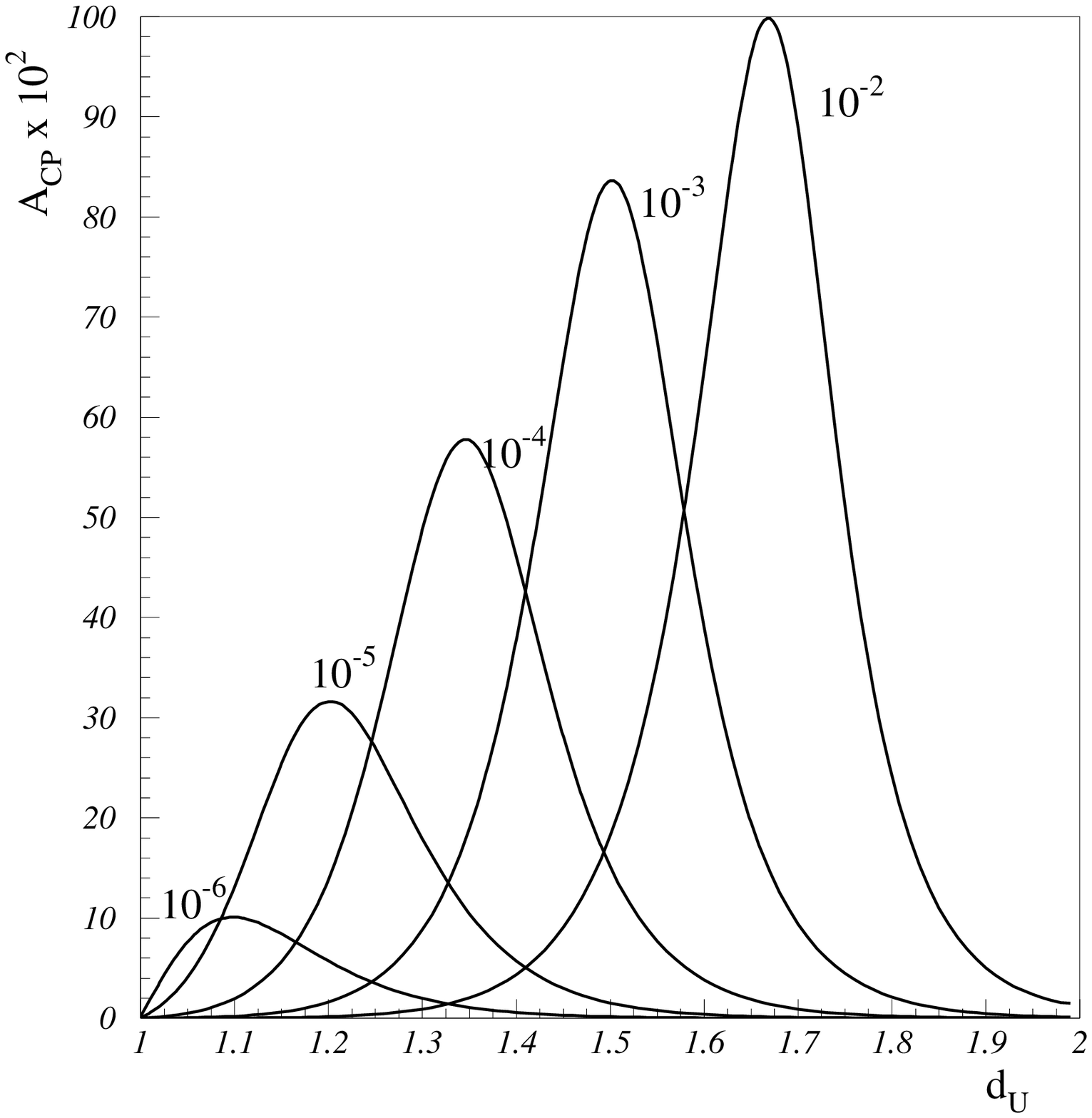}
\caption{ \label{longitvector} Longitudinal vector unparticle
contribution to Br and $A_{\rm CP}$ of $B^- \to \tau \bar{\nu}$.
From right to left, the corresponding coefficients are $C^{bu}_V
C^l_V = 10^{-2}$, $10^{-3}$, $10^{-4}$, $10^{-5}$, $10^{-6}$.}
\end{figure}

In Fig.(\ref{longitvector}), we study the contribution of
longitudinal vector unparticle to Br and $A_{\rm CP}$ of $B^\pm
\to \tau^\pm \nu(\bar{\nu})$, and include the constraint from the
Br when we consider the theoretical maximum of $A_{\rm CP}$.
We plot the Br and $A_{\rm CP}$ as a function of $d_{\cal U}$ with
different parameters of $C^q_V C^l_V = 10^{-2}$, $10^{-3}$,
$10^{-4}$, $10^{-5}$, $10^{-6}$. We find $A_{\rm CP}$ can be as
large as $100\%$ with $C^q_V C^l_V = 10^{-2}$, even under the
constraint of the Br. The reason is that r (see eq. (\ref{r}) for
its definition) can reach 1 due to the enhancement of
$(\Lambda/m_B)^2$ (see eq.(\ref{vec})). When $C^q_V C^l_V =
10^{-6}$, the maximum of $A_{\rm CP}$ can be $10\%$ under the
constraint of the Br. 

To see the possible effects of unparticles we should impose
constraints from precisely measured QED processes. The
contributions of charged unparticles to Bhabha scattering arise at
the loop level and consequently are small compared with those at
the tree level, such as the $\nu e$ scattering. Because there is
no data of $\nu\tau$ scattering we assume unparticle-leptons
coupling universality to proceed. The interaction in
eq.(\ref{lep}) contributes to $\bar{\nu}_e e \to \bar{\nu}_e e$
elastic scattering through s-channel charged unparticle
exchange\footnote{The contributions of SM singlet unparticles to
the scattering, which arise through t-channel unparticle exchange,
have been investigated in ref.~\cite{enu}.}. For the contribution
of scalar unparticle, the differential cross section is
straightforwardly calculated and the result is
\be
\label{enuenu}
\frac{d\sigma}{d T} &=& |\frac{A_{d_{\cal U}}}{2\sin(d_{\cal U} \pi)}|^2
  |C^l_S C^l_S|^2
  \frac{m_e^5}{2 \pi}
  \frac{s^{2d_{\cal U}-4}}{\Lambda^{4d_{\cal U}}} \, .
\ee
The contribution from vector unparticle can be written as \be
\frac{d\sigma}{d T} &=& |\frac{A_{d_{\cal U}}}{2\sin(d_{\cal U}
\pi)}|^2
  |C^l_V C^l_V|^2
  \frac{m_e}{\pi (s-m_e^2)^2} (s + t - m_e^2)^2
  \frac{s^{2d_{\cal U}-4}}{\Lambda^{4d_{\cal U}-4}} \, ,
\ee where the terms proportional to $m_e^5$ have been neglected.
In above two equations T is the energy of the recoil electron and
s, t, u are the Mandelstam variables, which are a function of
incoming neutrino energy $E_\nu$.

We now consider the constraint on unparticle-lepton couplings from
the TEXONO experiment searching for the neutrino magnetic
moments~\cite{Wong:2006nx}. In the case of scalar unparticle,
there are essentially no constraints on the parameter space,
because of the $m_e^5$ factor in the cross section of $\bar{\nu}_e
e \to \bar{\nu}_e e$ scattering in eq.(\ref{enuenu}). In the
vector unparticle exchange in $\bar{\nu}_e e$ scattering, $d_{\cal
U}$ are constrained to be larger than $1.5329$, $1.3882$,
$1.2453$, $1.1034$ for different $C^{bu}_V C^l_V = 10^{-2}$,
$10^{-3}$, $10^{-4}$, $10^{-5}$, respectively. When $C^{bu}_V
C^l_V = 10^{-6}$, the constraint on $d_{\cal U}$ is very small. We
find that $A_{\rm CP}$ can still reach their peak value shown in
Fig. (\ref{longitvector}) under the constraint.

If we further assume both unparticle-lepton coupling universality
and unparticle-quark coupling universality, the interactions in
eq.(\ref{lep}) and (\ref{qua}) also contribute to the $\pi^- \to
\mu \bar{\nu}_\mu$ decay channel. The constraint on unparticle
couplings arising from the decay is quit stringent. In the case of
scalar unparticle, $d_{\cal U}$ are constrained to be larger than
$1.4550$, $1.3882$, $1.3643$, $1.3045$ for different $C^{bu}_S
C^l_S = 1.$, $0.1$, $0.05$, $0.01$, respectively. When $C^{bu}_S
C^l_S = 1.$, $A_{\rm CP}$ is bound to be smaller than $8\%$, while
when $C^{bu}_S C^l_S = 0.1$, $0.05$, $0.01$, $A_{\rm CP}$ is
smaller than $2\%$. In the case of longitudinal vector unparticle,
the constraint from $Br(\pi^- \to \mu \bar{\nu}_\mu)$ is even more
stringent. $A_{\rm CP}$ is smaller than $1\%$, when $C^{bu}_V
C^l_V$ takes values as in Fig.(2) and $d_{\cal U} < 2$ .

We know from the table 1 in ref.~\cite{3799} that SuperB factories
will at best be able to measure the branching ratio of $B\to
\tau\nu$ at a $5 \%$ level, which would make a measurement of CP
asymmetry at $8\%$ almost impossible. However, from our
understanding on quarks so far, it is expected that the
universality of unparticle-quark coupling could be violated
significantly, in particular, for the third generation. Therefore,
the stringent constraint from the $Br(\pi^- \to \mu
\bar{\nu}_\mu)$ may not be applicable to the B decay $B^\pm\to
\tau \nu$ and consequently the CP asymmetry of $B\to\tau\nu$ can
be large enough to be measured, as shown above. In this case it is
also interesting to measure the direct CP violation in $B\to
\mu\nu$ since B factories have a better identification for $\mu$
than that for $\tau$. It is obvious from eqs.(14), (15) that
$A_{CP}$'s are the same for $l=e, \mu, \tau$ if the couplings
$C_{S,V}^l$ for $l=e, \mu, \tau$ are the same while the branching
ratio of $B\to\tau \nu$ has a enhancement of order of $100$ than
that for $B\to\mu\nu$ due to $m_{\tau}\doteq 17\;m_{\mu}$.

In summary, we have calculated the effects of unparticles in the
decays $B^\pm\to l^\pm \nu$. We have estimated the effects of
scale invariance breaking on Br and CP violation, which is due to
the interaction of electromagnetically charged unparticles with SM
particles, and the result is that effects are not significant in
general. It is found that the direct CP violation in $B\to l\nu$,
which is zero in SM and all other well-known models so far, can
reach $30\%$ for scalar unparticle contribution, and $100\%$ for
longitudinal vector unparticle contribution under the constraints
of ${\rm Br}(B\to\tau\nu)$ and $\nu e$ elastic scattering.
However, if the universality for both unparticle-lepton and
unparticle-quark couplings is assumed, the direct CP violation in
$B\to l\nu$ can only reach at most $8\%$ and $1\%$ for scalar and
vector unparticles respectively for $d_{\cal U} < 2$ due to the
constraint from the decay $\pi^- \to \mu \bar{\nu}_\mu$. From our
understanding on quarks so far, it is expected that the
universality of unparticle-quark coupling could be violated
significantly, in particular, for the third generation. Therefore,
the stringent constraint from the $Br(\pi^- \to \mu
\bar{\nu}_\mu)$ may be not applicable to the B decay $B^\pm\to
\tau \nu$. If the direct CP violation in $B\to l\nu$ is observed
in the future it would give strong evidence for the existence of
unparticle stuff.\\

{\bf \Large Acknowledgement}\\

The work was supported in part by the Natural Science
Foundation of China (NSFC), grant 10435040 and grant 90503002.
We'd like to acknowledge the comments
from Chunhui Chen, Kingman Cheung and Paul D. Jackson.


\end{document}